\documentclass[aps,prd,nofootinbib,twocolumn,floatfix,superscriptaddress,secnumarabic,reprint,showpacs]{revtex4}
\pdfoutput=1
\usepackage{amsfonts,amssymb,epsfig,verbatim,mathbbol,amstext,amsmath,psfrag}
\usepackage{graphicx}
\usepackage[    bookmarks,
                 bookmarksopen = true,
                 bookmarksnumbered = true,
                 linktocpage,
                 colorlinks = true,
                 linkcolor = blue,
                 urlcolor  = blue,
                 citecolor = blue,
                 anchorcolor = green,
                 hyperindex = true,
                 hyperfigures]
		{hyperref}
\usepackage[latin2]{inputenc}
\usepackage{t1enc}
\usepackage{amsmath}
\usepackage{subfigure}
\usepackage{dsfont}
\usepackage{slashed}
\usepackage{multirow}
\usepackage{lscape}
\usepackage{wrapfig}
\usepackage{array}
\usepackage{mciteplus}

\usepackage{setspace}

\usepackage{colordvi}
\usepackage{lscape}
\usepackage{rotfloat}
\usepackage{rotating}
\usepackage{color}

\usepackage{cmap}
\newcommand{\dd}{\textmd{d}}
\newcommand{\be}{\begin{equation}}
\newcommand{\ee}{\end{equation}}
\newcommand{\Tr}{\textmd{Tr}}
\newcommand{\Z}{\mathcal{Z}}
\renewcommand{\O}{\mathcal{O}}
\newcommand{\M}{\mathcal{M}}
\newcommand{\C}{\mathcal{C}}
\newcommand{\expv}[1]{\left \langle #1 \right \rangle}
\newcommand{\chib}{\chi}

\newcommand{\Li}{\textmd{Li}}

\long\def\symbolfootnote[#1]#2{\begingroup%
\def\thefootnote{\fnsymbol{footnote}}\footnote[#1]{#2}\endgroup}

\def\F{\mathcal{F}}

\newcommand{\Ds}{\slashed{D}}

\def\mpi{m_\pi}
\def\mupi{\mu_\pi}

\begin{document}
\title{Magnetic structure of isospin-asymmetric QCD matter in neutron stars}

\author{G.~Endr\H{o}di}
\affiliation{Institute for Theoretical Physics, Universit\"at Regensburg, D-93040 Regensburg, Germany\\ E-mail: \href{mailto:gergely.endrodi@physik.uni-r.de}{gergely.endrodi@physik.uni-r.de}}

\begin{abstract}
We study QCD under the influence of background magnetic fields and 
isospin chemical potentials using lattice simulations. 
This setup exhibits a sign problem which is circumvented 
using a Taylor-expansion in the magnetic field. The ground state of the system in the 
pion condensation phase is found to exhibit a pronounced diamagnetic response. We 
elaborate on how this 
diamagnetism may contribute to the pressure balance in the inner core of strongly 
magnetized neutron stars.
In addition we show that the onset of pion condensation shifts to larger chemical 
potentials due to the enhancement of the charged pion mass for growing magnetic fields.
Finally, we summarize the magnetic nature of QCD matter on the 
temperature-isospin chemical potential phase diagram.
\end{abstract}


\maketitle

\section{Introduction}

The elementary degrees of freedom of 
the high-energy phase of Quantum Chromodynamics (QCD) are
deconfined quarks and gluons. 
One physical environment where this deconfined phase may exist is the inner core of dense 
neutron stars
-- 
compact stellar objects 
created 
during the gravitational collapse of massive stars. 
A certain class of neutron stars (magnetars) exhibits intense magnetic 
fields of strengths up to $10^{14-15}\textmd{ G}$ at the star surface. 
These extreme magnetic fields are presumably 
generated by a convective dynamo mechanism during the first few seconds 
after the collapse~\cite{Duncan:1992hi},  
and are believed to be responsible for the strong electromagnetic activity of the 
star in the form of gamma-ray and X-ray bursts~\cite{Thompson:1996pe}. 
Magnetic fields can induce a strong deformation of the star, leading to the 
radiation of gravitational waves~\cite{Cutler:2002nw}. 
During mergers of binary neutron 
star systems, magnetic fields can even be amplified drastically~\cite{Price:2006fi} and 
have a significant impact on the emitted gravitational signal~\cite{Anderson:2008zp}. 
Thus, an understanding of the behavior of neutron star matter in magnetic fields 
is desired in many respects. 

While the magnetic fields at the surface of the star 
are typically determined using measurements of 
rotation periods and time derivatives thereof~\cite{Harding:2006qn}, 
the strength of the field in the deep interior of the star 
is highly uncertain. On general grounds 
the field $B$ is expected to become enhanced towards the star center, 
see, e.g., Ref.~\cite{Bocquet:1995je}.
The maximal possible strength is estimated to be  
$B\approx 10^{18}\textmd{ G}$ 
based on the equality of the gravitational and magnetic energies~\cite{Harding:2006qn}. 
These extreme values for $B$ are in the regime 
where a competition between the electromagnetic and the strong forces takes 
place and the magnetic properties of QCD matter as a medium become important.

An additional characteristic aspect of neutron stars is the isospin asymmetry that 
develops in their interior through protons converting into neutrons and neutrinos via 
electron capture. The core is thus described by a high baryonic density accompanied by a 
considerable isospin density. In the grand canonical approach to statistical physics, 
these densities are 
controlled by the corresponding chemical potentials. On the level of the (up and 
down) quark content, they are written as
\be
\mu_B=3(\mu_u+\mu_d)/2, \quad\quad\quad \mu_I=(\mu_u-\mu_d)/2.
\ee
To understand the structure of the dense and strongly magnetized inner core, 
an investigation of the combined effect of $\mu_f$ ($f=u,d$ labels 
the quark flavors) and $B$ on the ground state of QCD matter is necessary.
In the strongly interacting regime this is only 
possible using non-perturbative approaches like lattice simulations. 

To leading order in $B$, the magnetic response of the system 
is characterized by the magnetic susceptibility,
\be
\expv{\chib}=-\frac{1}{V} \left.\frac{\partial^2 \F}{\partial (eB)^2} \right|_{B=0},
\label{eq:chidef}
\ee
defined in terms of the free energy $\F=-T\log\Z$ of the system ($V$ denotes the 
three-dimensional volume, $T$ the temperature and $\Z$ the partition function). 
Here we considered the magnetic field 
in units of the elementary charge $e>0$. Note that the first derivative of $\F$ 
at $B=0$ vanishes due to parity symmetry.
The sign of $\chi$ distinguishes between paramagnetism ($\chi>0$) 
and diamagnetism ($\chi<0$). 

In this letter we calculate the magnetic susceptibility at nonzero isospin chemical potentials 
and discuss the structure of the $\mu_I-B-T$ phase diagram using numerical lattice QCD 
simulations. Our results indicate a strong 
diamagnetic response at high $\mu_I$, where charged pions condense. 
This response is related to the superconducting nature of the pion condensate.
We also present an argument suggesting that $\chi$ is less 
affected by the baryonic chemical potential and, thus, $\mu_I$ is the most relevant 
control parameter for the magnetic response of QCD at nonzero densities. 
In addition, we discuss 
the impact of the diamagnetic nature of the isospin-asymmetric state 
on magnetars. 
Assuming typical 
magnetic field configurations, this diamagnetism results in a considerable anisotropic 
force that can compete with the gravitational pressure in the core. This may have implications 
on, e.g., the convective processes in the interior of the star.

\section{Lattice setup and observables}
\label{sec:setup}

We consider QCD with two quark flavors, described by a 
two-component quark field $\psi=(\psi_u,\psi_d)$ and 
the fermion matrix
\be
M = 
\begin{pmatrix}
 \Ds(\mu_I,q_u)+m & \lambda \gamma_5\\
 -\lambda \gamma_5 & \Ds(-\mu_I,q_d)+m \\
\end{pmatrix}
\label{eq:fermmat}
\ee
where $q_f$ is the electric charge of the quark of flavor $f$, and we 
assumed a degenerate mass $m$.
Here, $\Ds=\gamma_\nu D_\nu$ is the Dirac operator with the $\mathrm{SU}(3)\times\mathrm{U}(1)$ covariant derivative $D_\nu=\partial_\nu+iA_\nu + iq_f A^{\rm em}_\nu$. The 
Abelian vector potential is chosen such that it generates a 
magnetic field in the $z$ direction, 
$A^{\rm em}_4=0$, $\nabla \times \mathbf{A}^{\rm em} = \mathbf{B} \parallel \mathbf{\hat{z}}$.

The term proportional to $\lambda$ is inserted in Eq.~(\ref{eq:fermmat}) as 
a small explicit breaking to allow the observation of pion condensation,
corresponding to the expectation value 
$\expv{\pi}\equiv\expv{\bar \psi_u \gamma_5 \psi_d - \bar \psi_d \gamma_5 \psi_u}$. 
This expectation value signals the spontaneous breakdown of isospin symmetry (more precisely: an
$\mathrm{U}(1)$ subgroup of the full isospin group that is left intact at $m\neq0$ and $\mu_I\neq0$).
In a finite volume this spontaneous breaking cannot occur 
without a small explicit breaking. 
The coefficient $\lambda$ will be extrapolated to zero at the end 
of the analysis. 
Incidentally, for $\mu_I\gtrsim m$ the smallest eigenvalue of the Hermitian operator $M^\dagger M$ 
(which is used in the simulation algorithm) equals 
$\lambda^2$, such that the system becomes ill-conditioned as $\lambda\to0$. 
Nevertheless, for the values necessary to perform this extrapolation (see 
Sec.~\ref{sec:result}) this numerical problem still turned out to be feasible. 

This theory is simulated on a symmetric $N^4$ lattice with $N=8$ such that the spatial volume 
equals $V=L^3=(8a)^3$ with $a$ being the lattice spacing.
To allow for a cross-check of the algorithm and of the simulation code, 
we use the same lattice discretization as Ref.~\cite{Kogut:2002zg}: the 
plaquette gauge action $S_p(\beta)$ and rooted staggered quarks. 
The partition function is obtained via the 
functional integral over the gluonic links $U_\nu=\exp(iaA_\nu)$ as
\be
\Z = \int \prod_\nu\mathcal{D} U_\nu \,e^{-S_p(\beta)} \det M^{1/4}.
\label{eq:partfunc}
\ee
The inverse gauge coupling 
is $\beta\equiv6/g^2=5.2$ and the quark mass in lattice units equals $ma=0.025$. 
The lattice spacing 
is determined using the Wilson flow~\cite{Luscher:2010iy} via the $w_0$ scale proposed in Ref.~\cite{Borsanyi:2012zs} to be $a=0.299(2)\textmd{ fm}$. 
The linear size of the system is $L\approx 2.4\textmd{ fm}$. 
Thus, $L^{-1}\approx 80 \textmd{ MeV}$ is well below the 
finite temperature deconfinement transition and the system may be approximated as being 
at zero temperature.
Through fitting the 
pseudoscalar propagator we get for the pion mass in lattice units $m_\pi a=0.402(5)$, 
giving $m_\pi\approx 260 \textmd{ MeV}$. 

Using the $\gamma_5$-hermiticity of the Dirac operator 
(for staggered quarks the role 
of $\gamma_5$ is played by $\eta_5=(-1)^{n_x+n_y+n_z+n_t}$), 
\be
\gamma_5 \Ds(\mu_I,q) \gamma_5 = \Ds(-\mu_I,q)^\dagger,
\ee
one can prove that the fermionic action $S_f\propto-\log \det M$ is real and positive 
if the electric charges of the two flavors coincide. 
However, having 
$q_d=-2q_u=-e/3$ is essential to capture the fact that the particles 
excited by the isospin chemical potential are the {\it charged} pions. 
Therefore, in the presence of the magnetic field, $S_f$ becomes complex and 
cannot be simulated using conventional lattice Monte-Carlo methods. 

We tackle this complex action problem by simulating at $\mu_I\neq0$ and 
$B=0$ -- where $S_f$ is real and positive -- and performing a (leading-order) Taylor-expansion in the 
magnetic field. This expansion involves the derivatives of the free energy 
with respect to $eB$ -- starting with $\chib$ of Eq.~(\ref{eq:chidef}). 
One technical complication is that on a finite lattice with periodic boundary 
conditions, the magnetic flux $\Phi=eB\cdot L^2$ is quantized~\cite{'tHooft:1979uj}, 
making the derivative with respect to $eB$ ill-defined. 
An advantageous strategy is to consider a 
modified magnetic field, for example one that is positive in one half and negative 
in the other half of the lattice~\cite{Levkova:2013qda}, 
$\mathbf{B} = B \,\textmd{sign}(L/2-x)\cdot \mathbf{\hat{z}}$. 
To implement this magnetic field configuration, the Abelian links $u^f_\nu=\exp(ia q_fA^{\rm em}_\nu)$ for the flavor $f$ are chosen as~\cite{Levkova:2013qda}
\be
\begin{split}
u^f_y(n_x)&=e^{ia^2q_f B \cdot(n_x-N/4)}, \quad\; n_x\le N/2,\\
u^f_y(n_x)&=e^{ia^2q_f B \cdot(3N/4 - n_x)}, \quad n_x>N/2,\\
u^f_\nu &=1, \hspace*{2.75cm}(\nu\neq y),
\end{split}
\ee
where $n_x=x/a$ denotes the $x$-coordinate of the sites.
The links (and, thus, also the magnetic field) satisfy periodic boundary conditions.
This setup was tested 
to give reliable results at $\mu_I=0$, where a direct simulation at $B\neq0$ is also 
possible~\cite{Bali:2014kia}. 
Nevertheless, we mention that finite volume effects may become enhanced due to 
the presence of the boundaries where the magnetic field changes sign, and for precision 
results our measurements should be repeated on larger volumes. 

Since the total magnetic flux is zero, the derivative with respect to $eB$ -- being a 
continuous variable -- can now be taken. Differentiating Eq.~(\ref{eq:partfunc}) twice, the susceptibility Eq.~(\ref{eq:chidef}) reads
\be
\expv{\chi} = \frac{\expv{\C_2}}{N^4}
,\quad\; \C_2 = \C_1^2+\C_1', \quad\; \C_1 = \frac{1}{4} \Tr(M^{-1}M'),
\ee
where $\expv{.}$ denotes the expectation value with respect to $\Z$ and the prime the derivative 
with respect to $eB$.
Besides the magnetic susceptibility, we also consider other observables like 
the chiral condensate, the pion condensate and the isospin density
\be
\expv{\bar\psi\psi} = -\frac{1}{V} \frac{\partial \F}{\partial m}, \quad
\expv{\pi} = -\frac{1}{V} \frac{\partial \F}{\partial \lambda}, \quad
\expv{n_I} = -\frac{1}{V} \frac{\partial \F}{\partial \mu_I} .
\label{eq:obs}
\ee
The second derivative of the expectation value of either of these observables $\O=\bar\psi\psi, \pi$ or $n_I$ can be found as
\be
\left.\frac{\partial^2 \expv{\O}}{\partial(eB)^2}\right|_{B=0} = 
\expv{\O''+2\O'\C_1 + \O\C_2} - \expv{\O}\expv{\C_2}.
\ee
In addition we will also discuss the Polyakov loop as a measure for deconfinement,
\be
P=\frac{1}{V} \sum_x \Tr \exp\left[\int \dd t A_4(x,t)\right].
\label{eq:Pldef}
\ee

We remark that the magnetic susceptibility 
contains an additive divergence at zero temperature and zero density, such that its renormalization reads
\be
\expv{\chib}^r = \expv{\chib} - \expv{\chib}_{\mu_I=T=0}.
\label{eq:chirdef}
\ee
This renormalization is related to the electric charge and wavefunction renormalization in 
QED~\cite{Schwinger:1951nm}. 
The $B$-dependent divergence cancels from the observables of Eqs.~(\ref{eq:obs}) and~(\ref{eq:Pldef}). 
For more details on this renormalization see, e.g., Ref.~\cite{Bali:2014kia}. 
In the following, the expectation value $\expv{.}$ is suppressed for the sake of brevity.

\section{Results}
\label{sec:result}

\begin{figure}[b]
\centering
\vspace*{-.1cm}
\includegraphics[width=8.0cm]{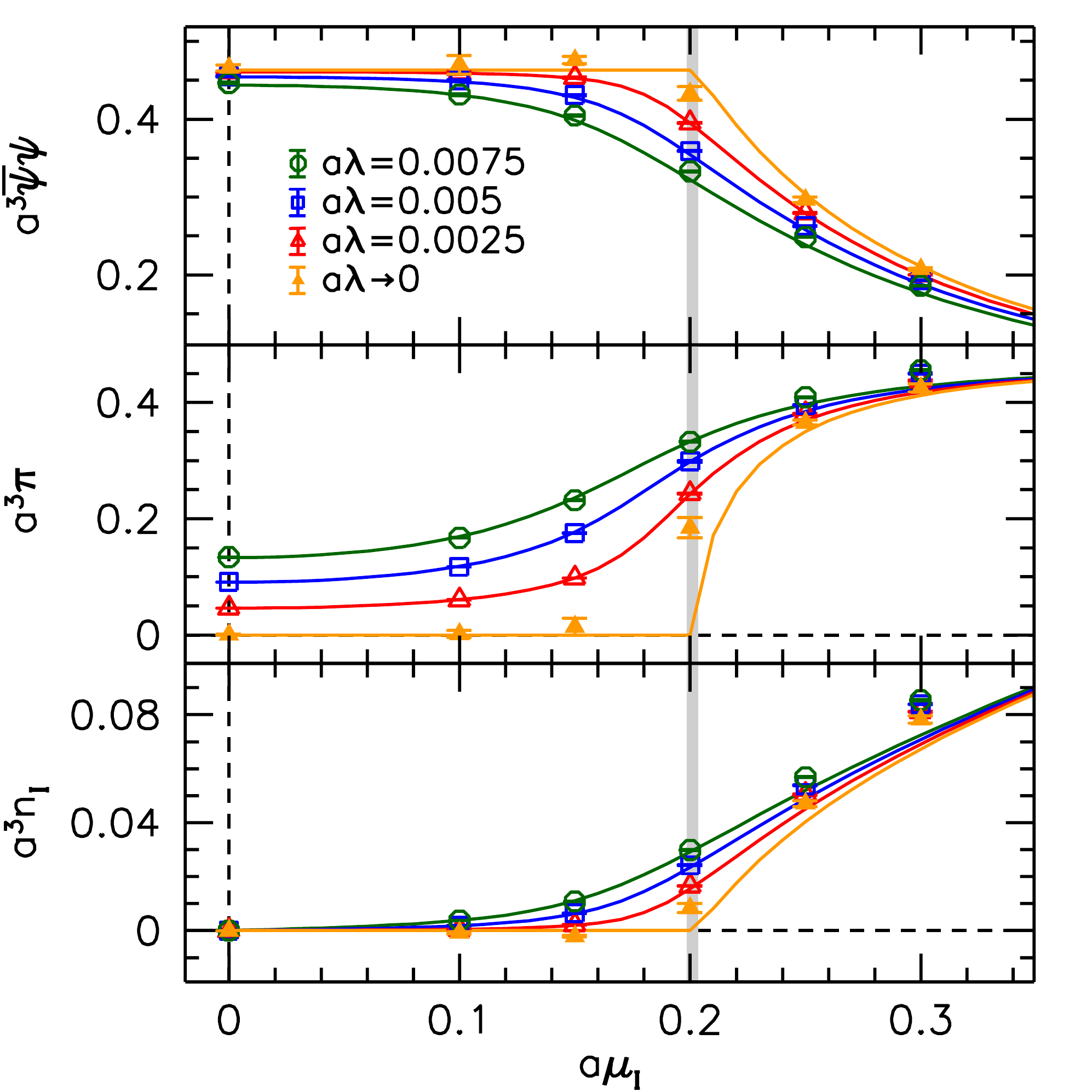}
\vspace*{-.1cm}
\caption{\label{fig:obs1}
Quark condensate (upper panel), pion condensate (middle panel) and 
isospin density (lower panel) as functions 
of the isospin chemical potential for various $\lambda$ values (red, blue and green points), 
a linear fit $\lambda\to0$ (yellow points), 
and a combined fit using $\chi$PT~\cite{Splittorff:2002xn} (solid lines). The onset 
of pion condensation at $m_\pi/2$ is indicated by the gray vertical line. 
}
\vspace*{-.1cm}
\end{figure}

We begin by considering the observables of Eq.~(\ref{eq:obs}) at vanishing 
magnetic field. Fig.~\ref{fig:obs1} shows the results at various values of 
the explicit symmetry breaking parameter $\lambda$ and a linear extrapolation to 
$\lambda=0$ for each $\mu_I$. The difference to a quadratic extrapolation 
is used to define the systematic uncertainty, and is included in the errors shown in the plot. 
We observe that the $\lambda=0$ limit of $\pi$ vanishes in the 
low-$\mu_I$ region, whereas it increases drastically for high chemical 
potentials, signalling the condensation of pions. 
The onset of this condensation is predicted by chiral 
perturbation theory ($\chi$PT) to occur at $\mu_I=m_\pi/2$~\cite{Son:2000xc}, 
in good agreement with the lattice data. 
All other observables are also insensitive to $\mu_I$ (in the $\lambda\to0$ limit) 
up to this onset value. This is in general referred to as the Silver Blaze phenomenon. 
The chiral condensate drops sharply above $\mu_I=m_\pi/2$, while the isospin 
density starts to grow at the condensation threshold. 
We note that our results are within statistical errors consistent with those 
of Ref.~\cite{Kogut:2002zg}, 
up to the fact that in Ref.~\cite{Kogut:2002zg} a different normalization convention 
was used. 
Our normalization is chosen such that the results at $\mu_I=0$ correspond to 
two degenerate flavors. We note moreover that lattice discretization effects 
start to dominate for $a\mu_I\gtrsim 1$ (not visible in the plot): here all lattice sites become 
occupied and the isospin density 
saturates at $a^3n_I^{\rm sat}=3/2$.

To perform the $\lambda$-extrapolation in a more effective manner, we 
consider $\chi$PT to describe the behavior of the observables 
for small $\lambda$ and for small $\mu_I$~\cite{Splittorff:2002xn}.
This dependence involves two parameters: 
the pion mass and the chiral condensate at $\mu_I=\lambda=0$, denoted by $G$, 
\be
\bar\psi\psi = G\cos \alpha, \quad
\pi = G \sin \alpha, \quad
n_{I} = \frac{ 4m G \mu_I }{m_\pi^2} \sin^2 \alpha,
\label{eq:chipt1}
\ee
where $\alpha$ is the vacuum angle, determined at the extremum of $\F$ by
\be
\sin ( \alpha - \phi) = \frac{4\mu_I^2}{m_\pi^2} \sin \alpha \cos \alpha, \quad
\phi = \arctan \lambda/m. 
\label{eq:chipt2}
\ee
We carry out a simultaneous fit of all three observables, using data points
for all values of $\lambda$, up to $a\mu_I=0.2$. Considering the pion mass as a free 
parameter of the fit, we obtain $am_\pi=0.4053(1)$, 
consistent with our previous determination using the pseudoscalar propagator. The latter 
is indicated by the gray line in the figure. 
For the condensate the fit gives $G=0.4657(2)$. 
In fact, $\chi$PT predicts pion condensation 
and chiral symmetry restoration to proceed simultaneously such that 
$\pi^2 + \bar\psi\psi^2$ remains constant. The lattice 
data do not support this prediction for $\mu_I\gtrsim m_\pi/2$, as $\pi$ turns out to be underestimated by 
$\chi$PT (this was also realized in Ref.~\cite{Kogut:2002zg}). 
However, below $m_\pi/2$ -- where the $\lambda$-dependence is most 
pronounced and, thus, the extrapolation cumbersome -- the $\chi$PT prediction is in excellent 
agreement with the lattice data. This comparison also reveals that the linear 
$\lambda\to0$ extrapolations of the lattice data using the available three 
$\lambda$ values -- with the exception of the points just at the condensation threshold -- 
are reliable. Note that a true phase transition only appears 
in the thermodynamic limit, and in a finite volume the observables slightly deviate from the 
behavior dictated by $\chi$PT around the onset chemical potential. 
As a side remark, we also mention that the structure of the chiral Lagrangian is the same  
for two-color QCD and for QCD with adjoint quarks. Thus, 
Eqs.~(\ref{eq:chipt1}) and~(\ref{eq:chipt2}) are also valid in these settings~\cite{Kogut:2000ek}.

\begin{figure}[b]
\centering
\vspace*{-.3cm}
\includegraphics[width=8.0cm]{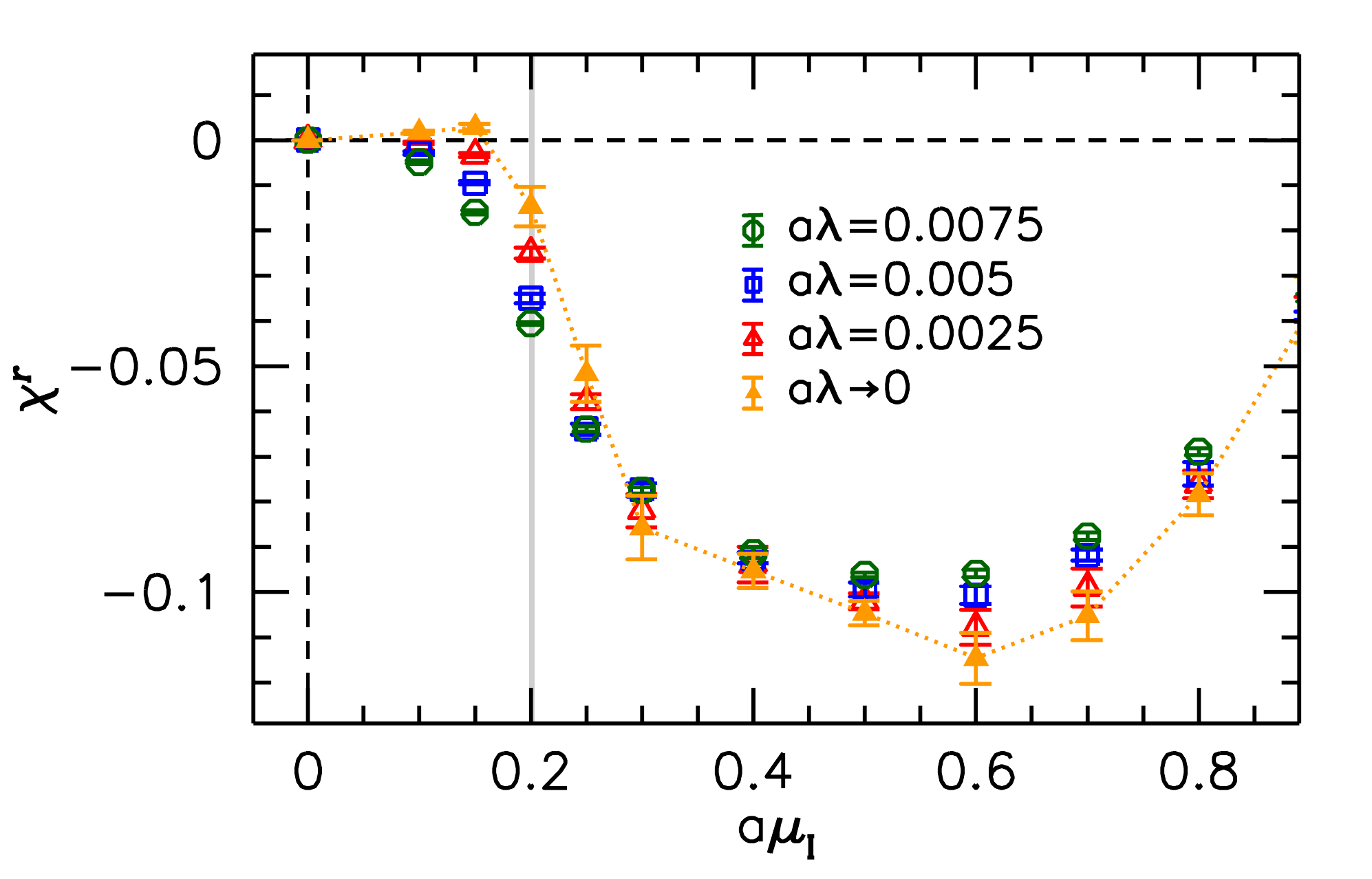}\\[-.1cm]
\includegraphics[width=8.0cm]{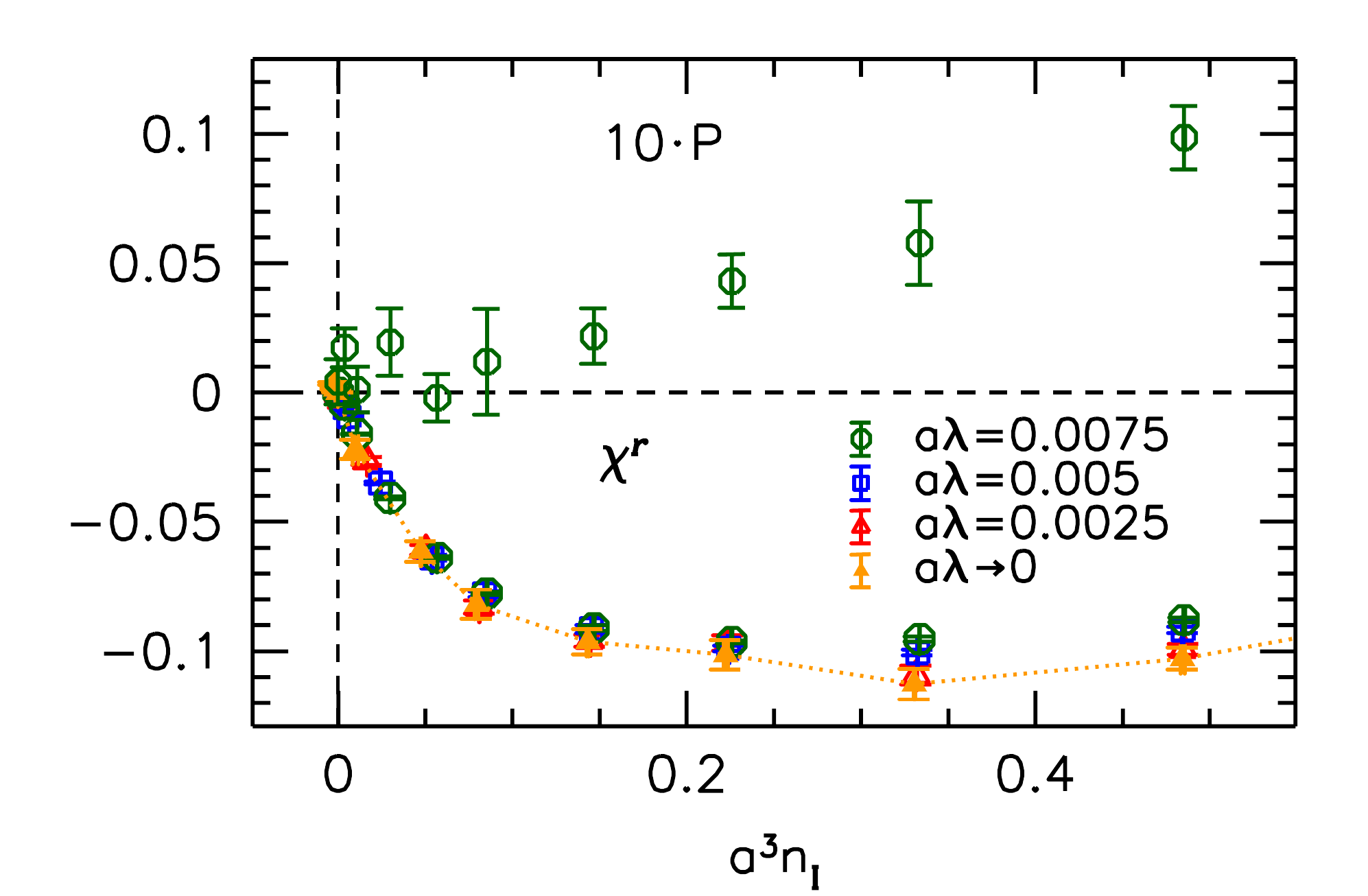}
\vspace*{-.2cm}
\caption{\label{fig:chi}
Renormalized magnetic susceptibility as function of $\mu_I$ (upper panel) and of $n_I$ (lower panel) 
for various values of $\lambda$ (red, blue and green points), 
and the $\lambda\to0$ extrapolations (yellow points), connected by the dotted line to guide the eye. 
The gray vertical line in the upper panel marks the onset of pion 
condensation. On the lower panel, the Polyakov loop for $a\lambda=0.0075$ is also included.
}
\vspace*{-.3cm}
\end{figure}

We proceed by performing a similar, linear $\lambda\to0$ extrapolation 
for the renormalized magnetic susceptibility Eq.~(\ref{eq:chirdef}). 
Subtracting the $\mu_I=0$ contribution at each $\lambda$ turned out to be advantageous 
here as it makes the $\lambda\to0$ extrapolation flatter. The results 
again show a Silver Blaze-type behavior up to $m_\pi/2$ and a rapid drop towards negative 
values beyond the onset of pion condensation (see upper panel of Fig.~\ref{fig:chi}). 
This implies that the QCD medium at low temperatures 
is {\it dia}magnetic for $\mu_I>m_\pi/2$. 
The diamagnetic response may be understood 
qualitatively from the fact that just above $m_\pi/2$, the system can be approximated as 
a dilute gas of 
pions. Pions are spinless and couple to the magnetic field only via their 
angular momentum. This coupling gives rise to a Landau-type diamagnetism, i.e.\ $\chi^r<0$.

In fact, the condensate of non-interacting point-like pions is expected to be superconducting 
and, as a result, a perfect diamagnet, which expels the magnetic field completely. 
This is demonstrated by explicit calculation in App.~\ref{sec:suscpion}. 
In full QCD, pions are not point-like free particles but interacting composite objects. 
This interaction poses an upper limit both on the density and on the conductivity.
As a result, the magnetic susceptibility remains below its free-case value.
In addition, unlike the chemical potential for free pions, 
the isospin chemical potential in QCD is not bounded by the pion mass. 
For $\mu_I>m_\pi/2$, the density and the susceptibility are again influenced 
predominantly by QCD interactions. 
In particular, the rise of the Polyakov loop, Eq.~(\ref{eq:Pldef}) -- as shown in 
the lower panel of Fig.~\ref{fig:chi} -- reveals that deconfinement and the enhancement of $n_I$ 
occur roughly simultaneously. This also implies that the pionic description breaks down. 
On the same figure, the susceptibility is also plotted as function 
of the isospin density, showing a 
linear section at low $n_I$ and a saturation to about $\chi^r=-0.1$ as the density increases.

For even higher chemical potentials, QCD asymptotic freedom allows to neglect the strong interactions 
completely. Then, the magnetic susceptibility can be calculated for free quarks, giving~\cite{Elmfors:1993bm}
\be
\chi^r \xrightarrow{\mu_I\to\infty} \frac{1}{4\pi^2} \sum_f (q_f/e)^2\cdot \log(\mu_I^2/\Lambda^2) > 0,
\label{eq:chi_asympt}
\ee
where the prefactor is related to the QED $\beta$-function 
and $\Lambda$ is a dimensionful scale (in the on-shell 
renormalization scheme of the free theory, $\Lambda=m$)~\cite{Bali:2014kia}.  
Thus, the susceptibility must eventually turn 
positive as $\mu_I$ increases.
To explore the region where $\chi^r$ crosses zero, 
further simulations on finer lattices are necessary.

\section{Interpretation}
\label{sec:interp}

\begin{figure}[t]
\centering
\includegraphics[width=7.5cm]{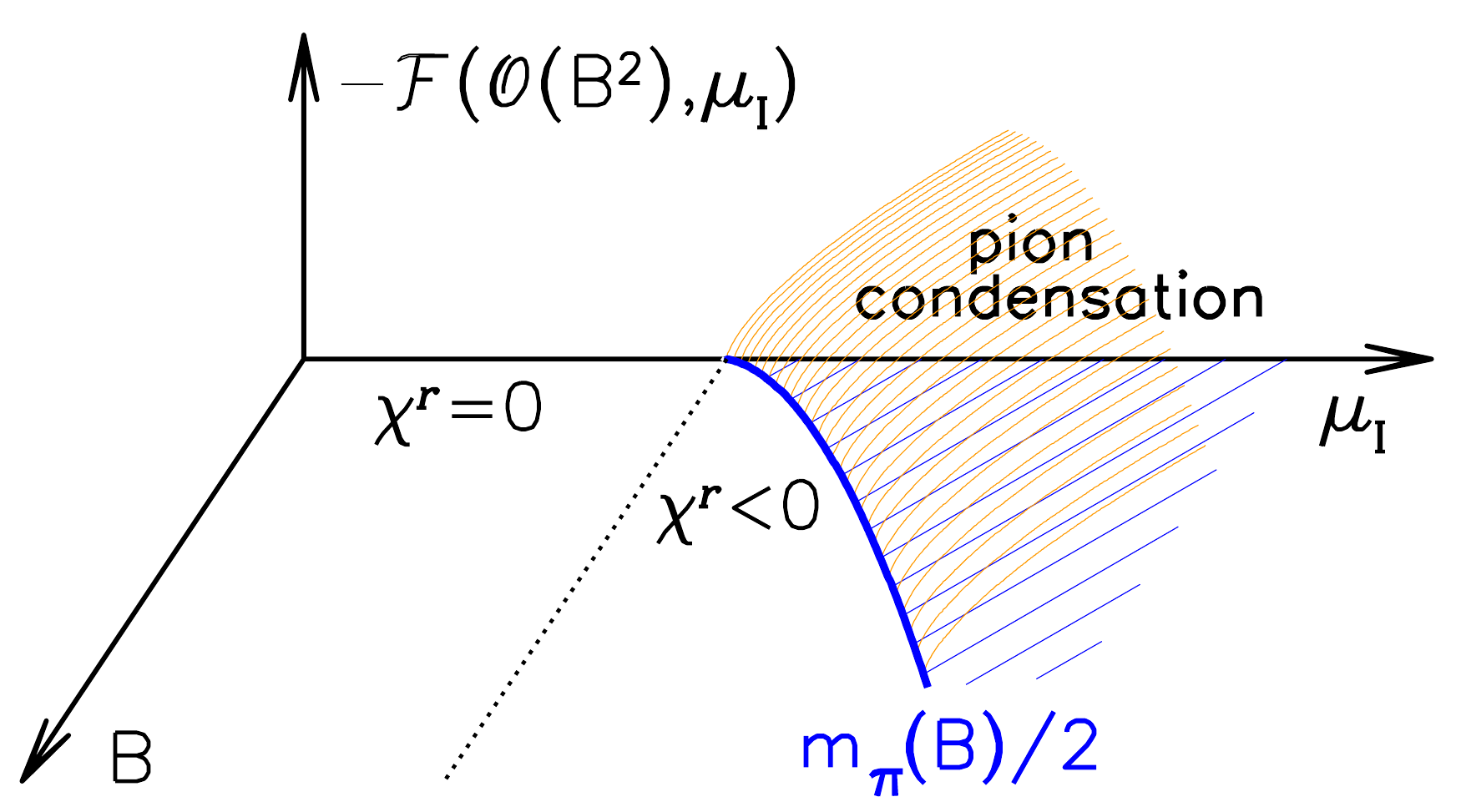}
\caption{\label{fig:F}
Illustration of the negative of the free energy as a 
function of $B$ and $\mu_I$ for small 
magnetic fields. The Silver Blaze 
region $\F=0$ expands as $B$ grows, implying a large negative 
magnetic susceptibility in the pion condensation phase.
}
\vspace*{-.1cm}
\end{figure}

Let us discuss the strong diamagnetic response above $\mu_I=m_\pi/2$ from a different 
point of view and consider how the charged pion mass responds to the magnetic field. 
Taking the pion as a point-like (relativistic) particle, the leading-order dependence reads
\be
m_{\pi}(B) = \sqrt{m_{\pi}^2(0)+eB},
\label{eq:mpi}
\ee
as a consequence of the lowest-Landau-level structure for a scalar particle. 
Note that Eq.~(\ref{eq:mpi}) is subject to corrections due to the $B$-dependence of 
the pion self-energy. These corrections are, however, small compared to the leading behavior~\cite{Colucci:2013zoa}. 
Indeed, recent lattice QCD results~\cite{Bali:2011qj} have confirmed Eq.~(\ref{eq:mpi}) up to $eB\approx 0.4 \textmd{ GeV}^2$. 
At $T=0$, the renormalized free energy can be calculated as the integral of the isospin density,
\be
-\frac{\F(B,\mu_I)}{V} = \int_0^{\mu_I} \!\dd \mu_I' \,n_I(B,\mu_I').
\ee
For $B=0$, taking into account the dependence $n_I(\mu_I)$ from Fig.~\ref{fig:obs1} 
(for $\lambda\to0$), 
this implies that $-\F$ is zero up to $m_\pi/2$ and becomes positive 
above the threshold. 
Let us now switch on a weak magnetic field and consider 
the free energy up to $\mathcal{O}(B^2)$. 
To this order, $\F(B,\mu_I)$ still vanishes for chemical potentials up to the 
corresponding pion mass. However, due to Eq.~(\ref{eq:mpi}), the 
Silver Blaze region expands as $B$ grows. This can only be maintained 
if the surface $-\F$ has a large negative curvature in the $B$-direction, i.e., 
through a large negative susceptibility 
(for an illustration see Fig.~\ref{fig:F}). Thus, the pion condensation phase must exhibit 
strong diamagnetism -- in line with our results presented in Sec.~\ref{sec:result}. 

\begin{figure}[b]
\centering
\vspace*{-.4cm}
\includegraphics[width=8.0cm]{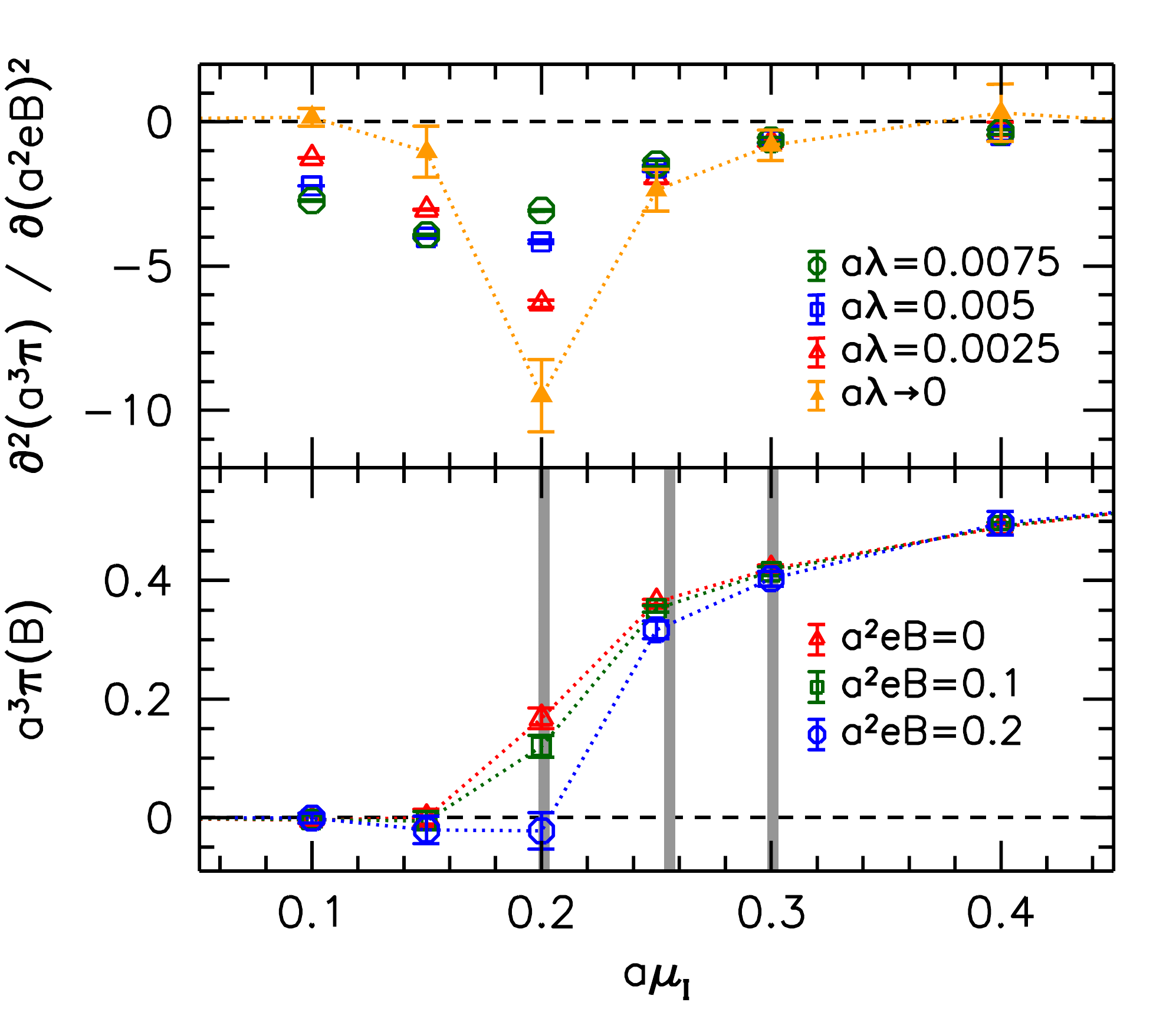}
\vspace*{-.2cm}
\caption{\label{fig:pi_B}
Second derivative of the pion condensate (upper panel; the symbols are the same 
as in Fig.\protect~\ref{fig:chi}) 
and the $\lambda\to0$ limit of the pion condensate at nonzero magnetic fields 
using the leading Taylor-expansion (lower panel). The gray vertical 
lines indicate the pion mass for each magnetic field (increasing from left to right).
}
\vspace*{-.1cm}
\end{figure}

Note that the key component in the above argumentation was the scalar nature of the pion, 
which allowed for a phase with Bose-Einstein condensation and, at the same time, implied an increase in 
the mass as $B$ grows, Eq.~(\ref{eq:mpi}). 
For the baryonic chemical potential $\mu_B$, the excited particles are protons and neutrons. In this 
case the baryon density grows much slower $n_B\propto(\mu_B^2-m_B^2)^{3/2}$, and condensation 
can only occur via Cooper-pairing if the ground state is in the superfluid phase. 
Moreover, in the spin-half channel, 
the mass (to leading order) is independent of the magnetic field (both for neutrons and for protons), 
implying that the Silver Blaze region is also insensitive to $B$. Thus, $\chi^r$ is expected to be 
suppressed for $\mu_B$ just above $m_B$. 
Note that in nature nucleons have anomalous magnetic moments that 
induce a dependence of the mass on $B$, and through that 
a nonzero value for $\chi^r$, 
but due to the absence of direct condensation and due to the larger mass, 
these effects are not expected to 
produce a pronounced behavior like the one seen in Fig.~\ref{fig:chi}.

To back up the picture described above, we also determined the lowest-order expansion 
coefficients of the observables in Eq.~(\ref{eq:obs}). 
In Fig.~\ref{fig:pi_B} we show the second derivative of the 
pion condensate with respect to $eB$ and the reconstructed observable
$\pi(B)$ for a few values of the magnetic field. 
As $\lambda\to0$, the derivative is found to exhibit a pronounced dip around 
$\mu_I=m_\pi/2$ and, as a consequence, the rise in $\pi$ 
is shifted to higher isospin chemical potentials as $B$ is increased. 
The lattice data for the scalar condensate and for the isospin density 
show similar trends. 
Therefore, the results are in qualitative agreement with the discussion above, 
namely that the magnetic field shifts the onset of pion condensation 
to higher isospin chemical potentials. 
On the quantitative level, the results in the lower panel of Fig.~\ref{fig:pi_B}
suggest that this shift is less pronounced 
than the expectation based on Eq.~(\ref{eq:mpi}). 
(Note that the Taylor-expansion in $B$ breaks down at the 
phase transition, where $\F$ is non-analytic. Still, the reconstruction 
of $\pi(B)$ is expected to converge outside of the close vicinity of the onset isospin chemical potential.)

\section{Implication for magnetars}
\label{sec:nstar}

Next, we consider a possible implication of the diamagnetic pion condensed 
phase on the physics of strongly magnetized neutron stars. 
Charged pion condensation in neutron star cores has been the subject of discussion for a long 
time~\cite{Migdal:1990vm}, as it is expected to have significant implications 
for, e.g., the equation of state~\cite{Suh:2000ni} as well as neutron star cooling rates~\cite{Maxwell:1977zz}. 
Assuming charge neutrality, together with equilibrium for neutron $\beta$-decay and for the 
process $n\to p+\pi^{-}$, the threshold density for pion condensation was found to 
be at a few times nuclear matter density (see, e.g., Ref.~\cite{Maxwell:1977zz,Suh:2000ni}). 
The possibility of charged pion condensation and its consequences for neutron 
star physics have also been discussed more recently in, e.g., Refs.~\cite{Takatsuka:1993pw,Takahashi:2002ig,Ohnishi:2008ng}. 

The most probable constituents of the pion condensed core are neutrons, protons, 
negatively charged pions, electrons and muons~\cite{glendenning2000compact}. 
As we argued in Sec.~\ref{sec:interp}, pions are strongly diamagnetic, whereas the contribution 
of protons and neutrons to $\chi^r$ is expected to be much smaller in magnitude. 
We remark that for the electron -- in contrast to the other particles -- the typical neutron star core 
magnetic fields exceed 
the rest mass squared (in fact, by several orders of magnitude). Thus, for the effect considered 
below, the electron contribution to the susceptibility is to be calculated in the 
strong field regime and not at $B=0$. 
Another comment about the electron contribution is in order here. 
The free energy in the presence of magnetic fields contains a thermal/dense contribution at 
nonzero $T$ and/or $\mu$, as well as a vacuum term that stems from virtual particles at $T=\mu=0$ 
and that dominates in the strong field regime $eB\gg T^2,\mu^2,m^2$~\cite{Elmfors:1993bm}. 
However, as we will see, the vacuum term has no effect on the mechanism discussed below. 
The thermal/dense contribution to $\chi^r$ has been calculated several times in the literature
~\cite{Blandford:1982,Elmfors:1993bm,Elmfors:1994mq,Dong:2013hta} 
and was always found to be below a few percents 
for the magnetic field strengths considered here $B\approx 10^{18} \textmd{ G}$. 
Similarly small estimates for the muon, proton and neutron contributions
were also given in Ref.~\cite{Dong:2013hta}. 
Altogether we can thus estimate the total magnetic susceptibility in the magnetar core 
by the pionic contribution and take $\chi^r\approx -0.1$ as a typical value 
that we obtained. 

\begin{figure}[b]
\centering
\vspace*{-.2cm}
\includegraphics[width=8.0cm]{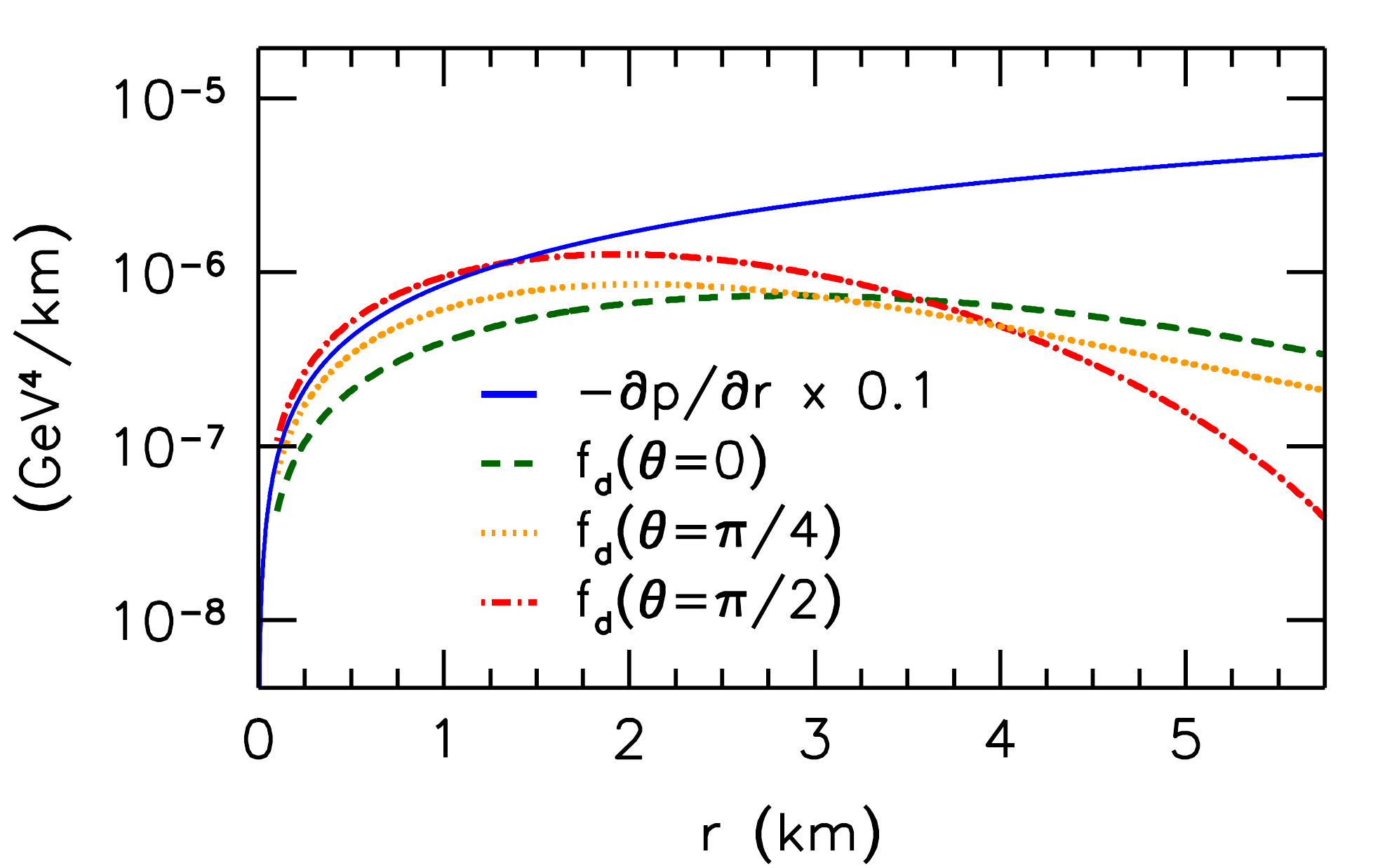}
\caption{\label{fig:comp}
The force generated by QCD diamagnetism 
(along various directions specified by the polar angle $\theta$) and 10$\%$ of the pressure gradient as 
functions of the radial coordinate.}
\vspace*{-.2cm}
\end{figure}

The proposed mechanism involves a gradient force that emerges in inhomogeneous 
magnetic fields. Namely, the minimization of the free energy induces the force density, 
\be
f_d = -\frac{1}{V} \nabla \F = \chi^r |eB| \,\nabla |eB|.
\label{eq:fd}
\ee
Note that this force is only sensitive to the magnetic properties of the medium, and 
neither the energy $B^2/2$ of the magnetic field, nor the above mentioned vacuum contribution in $\F$ 
contributes to $f_d$ 
(assuming that $B$ is constant in time, i.e.\ there is no feedback from matter to 
the magnetic field configuration). 
Note moreover that since the free energy is a Lorentz-scalar, $f_d$ only depends on the 
magnitude of $B$. 
We adopt the 
poloidal magnetic field profile $\mathbf{B}(r,\theta)$ 
of Ref.~\cite{Bocquet:1995je} for a rotating magnetar with 
radius $R=10\textmd{ km}$ and central field strength $1.5\cdot 10^{18} \textmd{ G}$. 
The rotation axis is given by $\theta=0$.
Inserting $\chi^r\approx -0.1$ for the susceptibility we obtain $f_d(r)$, see the 
curves in Fig.~\ref{fig:comp} for three fixed values of the polar angle $\theta$. 

The so obtained force density is to be compared to the gradient of the isotropic pressure profile $p(r)$ 
in the star. 
For a first approximation, we take the simplified case of a star with constant density
(see Eq.~(3.157) of Ref.~\cite{glendenning2000compact}), 
and consider typical values for the central pressure 
$p_c \approx 10^{34} \textmd{ Pa}$. 
The resulting gradient is also included in Fig.~\ref{fig:comp}, 
indicating that in this case the diamagnetic effect amounts to up to $10\%$ of the gravitational 
pressure gradient in the 
inner core $r\lesssim 3 \textmd{ km}$. The curves for $f_d(r,\theta)$ also reveal that the 
diamagnetic force is anisotropic and tends to push material from the center 
towards the equator. 
Thus, we expect that the diamagnetism of isospin-asymmetric QCD matter plays 
a relevant role for the 
description of convective processes in the inner core.
We mention that a similar mechanism in the case of heavy-ion collisions was 
discussed in Ref.~\cite{Bali:2013owa}.

\section{Conclusions}

We have discussed the QCD phase diagram in the $\mu_I-B$ plane for the first 
time using lattice simulations. This setup has a complex action problem, which was 
circumvented through a Taylor-expansion in $B$ at nonzero isospin chemical potentials. 
We measured thermodynamic observables for a wide range of $\mu_I$ values, 
in the Silver Blaze region, through the onset of pion condensation at $\mu_I=m_\pi/2$, 
up to lattice saturation. 
The results indicate that the condensation threshold is shifted to higher values of 
$\mu_I$ as $B$ grows, in qualitative agreement with the dependence 
$m_\pi(B)$ of the pion mass on the magnetic field.
We demonstrated how this tendency explains the observed strong diamagnetic behavior of the system 
in the pion condensation phase. The diamagnetic nature of the pion condensate is also 
predicted by free-case arguments, see the analytic calculation of the pionic susceptibility 
in App.~\ref{sec:suscpion}.

In addition, we also presented an argument suggesting that the magnetic response 
of the QCD ground state is most sensitive to isospin chemical potentials, and 
the baryon chemical potential is not expected to play a dominant role in this respect.
Our results were obtained on coarse lattices with a larger-than-physical pion mass, and thus should 
be considered exploratory. However, since the findings are 
understood in terms of general arguments like the existence of a pion condensation phase 
and the diamagnetic nature of pions, the results are not expected to 
change qualitatively if the physical point and the continuum limit are approached.

We conclude by sketching the magnetic nature of QCD matter on 
the $T-\mu_I$ phase diagram. At $\mu_I=0$, lattice simulations have shown that 
the susceptibility is positive for 
$T\gtrsim 120 \textmd{ MeV}$~\cite{Bonati:2013lca,Levkova:2013qda,Bali:2013owa,Bonati:2013vba,Bali:2014kia}, 
whereas it becomes slightly negative for lower temperatures~\cite{Bali:2014kia}. 
The latter diamagnetic region is predicted by 
$\chi$PT and by the Hadron Resonance Gas model~\cite{Bonati:2013vba,Bali:2014kia}, 
and also stems from the presence of charged pions.
However, while at $T=0$ and $\mu_I>m_\pi/2$, pions are created in abundance, 
at $T>0$ they are induced merely by thermal fluctuations. Accordingly, 
the diamagnetic response at $0<T\lesssim 120 \textmd{ MeV}$ is much weaker 
($\chi^r\approx-0.002$~\cite{Bali:2014kia})
than the one in the pion condensation phase ($\chi^r\approx -0.1$). 
Note also that around the deconfinement temperature
($T_c\approx 150 \textmd{ MeV}$) at $\mu_I=0$, 
the susceptibility was found to be significantly smaller ($\chi^r\approx 0.01$~\cite{Bali:2014kia}) 
than the magnitude of our results in the pion condensed phase. 
Well above the deconfinement transition temperature at $\mu_I=0$, 
the dominant degrees of freedom are quarks, giving rise to strong paramagnetism, with 
$\chi^r\propto \log(T)$, similarly as in Eq.~(\ref{eq:chi_asympt}), just with $\mu_I$ 
replaced by $T$~\cite{Bali:2014kia}. Thus, asymptotic freedom in QCD ensures that 
QCD matter is paramagnetic for very high values of $T$ and/or $\mu_I$.

\begin{figure}[t]
\centering
\includegraphics[width=8.cm]{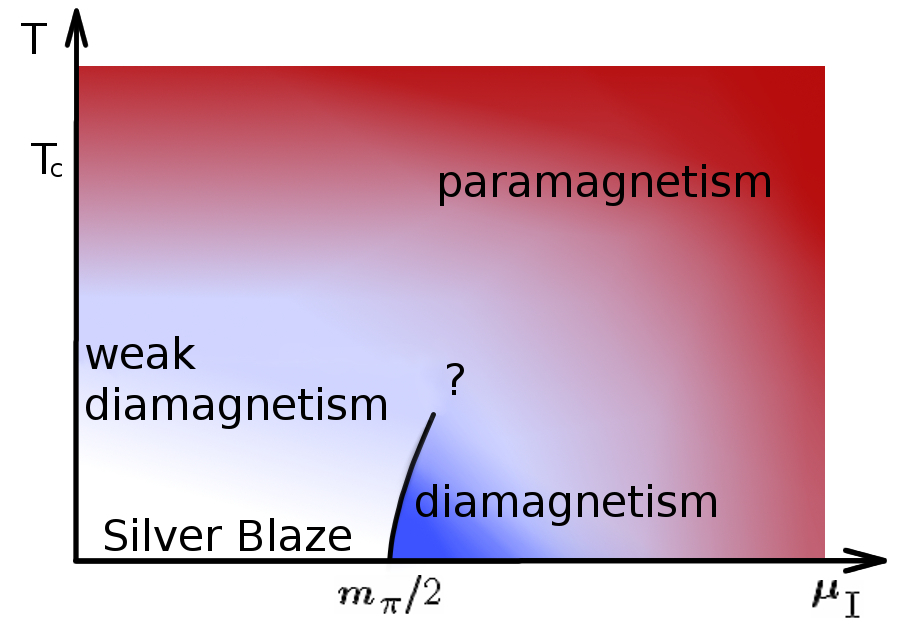}
\caption{\label{fig:conjecture}
Conjecture of the magnetic structure of the phase diagram in the $T-\mu_I$ plane. 
Regions with positive (negative) susceptibilities are represented by 
red (blue) and darker colors mark larger magnitudes. 
At $\mu_I=0$ the transition from diamagnetism to paramagnetism 
occurs slightly below the chiral crossover temperature 
$T_c\approx 150 \textmd{ MeV}$~\protect\cite{Bali:2014kia}.
The solid line at high $\mu_I$ represents a true phase transition separating 
the vacuum and the pion condensation phase. There 
may be additional phase transition lines in the interior of the diagram, indicated 
by the question mark.}
\end{figure}

Based on this picture, our conjecture of the magnetic phase diagram -- 
summarizing the dia- and paramagnetic regions of QCD matter -- is 
shown in Fig.~\ref{fig:conjecture}. 
The phase transition at $\mu_I=m_\pi/2$ is expected to bend 
to the right~\cite{Son:2000xc} and either persist towards higher temperatures 
or end at the deconfinement phase transition line if the latter exists. 
Which scenario is the case and 
whether there are additional phase transition lines in the diagram is
presently unclear. 
The isospin density has been shown to change the nature of the 
chiral transition in 8-flavor QCD~\cite{deForcrand:2007uz}. 
The structure of the $T-\mu_I$ phase diagram has been studied in 
various model frameworks as well~\cite{Zhang:2006gu,Sasaki:2010jz,Kamikado:2012bt,Stiele:2013pma}. 
A possible critical endpoint at nonzero isospin densities and 
magnetic fields was also discussed in Refs.~\cite{Kang:2013bea,Costa:2013zca} 
in model setups.
To determine the detailed structure of the interior of the phase diagram 
on the lattice for the interesting case of 2 or 2+1 flavors, 
further simulations are necessary.\\

\noindent {\bf Acknowledgements.} 
This work was supported by the DFG (SFB/TRR 55) and by the 
Alexander von Humboldt Foundation. 
The author thanks Gert Aarts, Massimo D'Elia, Kim Splittorff and Laurence Yaffe for useful discussions, and 
Gunnar Bali, Falk Bruckmann, Andreas Sch\"{a}fer for 
a careful reading of the manuscript. 

\appendix

\section{Susceptibility of free pions}
\label{sec:suscpion}

In this appendix we calculate the magnetic susceptibility of free 
non-relativistic pions. 
Since we aim to describe the condensation phase at low temperatures, 
where only one species -- say, the $\pi^-$ -- contributes, we consider 
a pion with charge $q=-e$ and exclude its positively charged 
antiparticle for simplicity. 
The energy levels read
\be
E_n=\mpi+\frac{|qB|}{2\mpi}(2n+1)+\frac{p_z^2}{2\mpi},
\ee
where $p_z$ is the momentum parallel to $B$, 
and we included the rest mass $\mpi$ of the pion in the energy. 
The matter contribution to the 
free energy of the pion gas
at temperature $T$ and chemical potential $\mupi$ in a finite volume $V$ is written as
\be
\F^m = \frac{|qB|\cdot V}{4\pi^2} \int\! \dd p_z \sum_n T\log\left[1-e^{-(E_n-\mupi)/T}\right].
\label{eq:Fmpion}
\ee
This expression does not contain the vacuum contribution, which stems from 
virtual pions at $\mupi=T=0$ in the presence of the magnetic field 
(see, e.g., Refs.~\cite{Elmfors:1993bm,Endrodi:2013cs}). 
However, the vacuum part does not contribute to the renormalized 
susceptibility at $B=0$, nor does it have an effect on the mechanism discussed in Sec.~\ref{sec:nstar}. It is therefore neglected in the following. 

The renormalized susceptibility is obtained as the second derivative of 
$\F^m$ with respect to $eB$. Since the sum and the integral in $\F^m$ are 
ultraviolet finite, these can be interchanged with 
the derivative,
\be
\chi^r = \frac{1}{4\pi^2} \int\! \dd p_z \sum_n \frac{(2n+1)/\mpi}{1-\exp\big(\frac{p_z^2}{2\mpi T}-\frac{\mupi-\mpi}{T}\big)}.
\ee
The sum over $n$ is calculated using $\zeta$-function regularization, while the 
integral gives a polylogarithm function,
\be
\chi^r = \frac{-1}{12\pi^2}\frac{\sqrt{2\pi T}}{\sqrt{\mpi}} \cdot \Li_{1/2}\big[e^{(\mupi-\mpi)/T}\big].
\label{eq:chirpi}
\ee
Note that for our bosonic system, the chemical potential cannot exceed $\mpi$. In the 
Silver Blaze region $\mupi<\mpi$, the 
zero-temperature limit of the polylogarithm function vanishes, resulting in $\chi^r=0$.
However, for $\mupi\to \mpi$ the susceptibility diverges for any infinitesimally small temperature,
\be
\lim_{T\to0}\chi^r(\mupi<\mpi)=0, \quad\quad \lim_{\mupi\to \mpi}\chi^r(T>0)=-\infty.
\ee
We remark that if the condensation phase is approached by gradually lowering 
the temperature at fixed density $n$, pions start to condense at 
the critical temperature~\cite{kapusta2006finite}
\be
T_c = \frac{2\pi}{\mpi} \left(\frac{n}{\zeta(3/2)}\right)^{2/3},
\ee
which is always non-vanishing. 

A remark about $\chi^r$ approaching $-\infty$ is in order.
Notice that $B$ equals the magnetic field acting in the medium, which 
is to be distinguished from the 
external magnetic field $H$ that would be present in the absence of pions. 
The two fields are connected by the magnetization,
\be
B=H+\M e, \quad\quad \M=-\frac{1}{V}\frac{\partial \F^m}{\partial (eB)}.
\ee
For weak fields, $\M=\chi^r\cdot (eB)$, which allows to express the 
magnetic permeability $p_m$ of the medium as
\be
p_m \equiv \frac{B}{H} = \frac{1}{1-e^2\chi^r}.
\ee
Clearly, for $\chi^r\ll1$ the difference between $B$ and $H$ is negligible 
and $p_m$ is very close to unity. 
However, for $\chi^r\to-\infty$, the permeability vanishes, signalling that the magnetic 
field is expelled from the system completely. 
This perfect diamagnetism is characteristic for superconductors. 
Indeed, at $\mupi=\mpi$ the system becomes superconducting due to the 
condensation of charged pions. 

\bibliographystyle{JHEP}
\bibliography{magnetar}

\end{document}